\documentclass[reqno,12pt]{article}
\usepackage{amsmath,amsfonts,amssymb,amsthm,amstext,amscd,eucal,xcolor}
\usepackage{color}
\usepackage[all]{xy}
\usepackage{mathtools}
\newcommand{\mathsym}[1]{{}} 
\usepackage[colorlinks=true,
            linkcolor=blue,
            urlcolor=blue,
            citecolor=blue]{hyperref}
\usepackage{enumerate}

\usepackage{graphicx}% Include figure files
\usepackage{bm}% bold math
\usepackage{cite,hyperref}
\usepackage{mathtools}
\usepackage[active]{srcltx}
\makeatletter \@addtoreset{equation}{section}

\makeatletter\renewcommand\section{\@startsection {section}{1}{\z@}%
                                   {-3.5ex \@plus -1ex \@minus -.2ex}%nn
                                   {2.3ex \@plus.2ex}%
                                   {\normalfont\large\bfseries}}
\renewcommand\subsection{\@startsection{subsection}{2}{\z@}%
                                     {-3.25ex\@plus -1ex \@minus -.2ex}%
                                     {1.5ex \@plus .2ex}%
                                     {\normalfont\bfseries}}

\DeclareMathAlphabet{\mathcal}{OMS}{cmsy}{b}{n}

\makeatother

\newcommand{\email}[1]{\footnote{E-mail: \href{mailto:#1}{#1}}}

\begin{document}

\title{  Vacuum Radiation in $z=2$ Lifshitz QED}
\author{R. Bufalo\email{rodrigo.bufalo@ufla.br} ~and T. Cardoso e Bufalo\email{taticardoso@ufla.br} \\
%EndAName
\textit{$^{1}$ \small Departamento de F\'isica, Universidade Federal de Lavras,}\\
\textit{ \small Caixa Postal 3037, 37200-900 Lavras, MG, Brazil}\\
\\
}

\maketitle
\date{}

\begin{abstract}
We discuss in this paper the vacuum Cherenkov radiation in the $z=2$ Lifshitz electrodynamics.
The improved ultraviolet behavior, in terms of higher spatial derivatives, and the renormalizable couplings, due to the time-space anisotropic scaling, present in the Lifshitz setting are extremely important in fulfilling the physical constraints in order to this vacuum process to happen.
We evaluate in details the instantaneous rate of energy loss for a charge, and also analyze the emission of very soft photons in this framework.
\end{abstract}

\section{Introduction}

In the past decade we have witnessed an important era of precision experiments in both particle physics and astrophysics, which has deepened our understanding of the Standard Model of particle physics (SM)
In addition to the experimental verification of several theoretical mechanisms and predictions of SM, physical phenomena that are not adequately explained by SM, the so-called Physics Beyond the Standard Model, have received attention and been scrutinized to great accuracy \cite{Lykken:2010mc}.
From the many possibilities to make contact with phenomena related to physics beyond the standard model, the most compelling ones are those proposals of violation of exact symmetries in field theories, that aim to make contact with Planck-scale physics.
In particular, models involving Lorentz violation have reached an important milestone in recent years due to systematic development and subjection to precision tests \cite{ref53,AmelinoCamelia:2008qg}.

Anomalous decay processes are valuable probes in the study of departures from Lorentz symmetry.
The main interest in decay processes is that they are affected in unexpected ways by Lorentz violation, meaning that forbidden processes can occur in certain regions of the parameter space \cite{ref53,Jacobson:2005bg}.
In the context of anomalous decays, highly energetic particles are the most interesting candidates to examine Lorentz violation because they are usually subject to instabilities;
in particular, instabilities involving photons in vacuum have caught interest in recent years because sufficiently energetic photon (usually from gamma-ray bursts) may decay as a manifestation of Lorentz  violation \cite{Ellis:2005wr,Heeck:2013cfa,Bonetti:2017pym}.
A widely explored anomalous process in the framework of Lorentz violating photons is the emission of vacuum Cherenkov radiation, an extremely important energy loss process for high-energy particles \cite{Lehnert,Kaufhold:2005vj,Kaufhold:2007qd}.

It is well known that ordinary Cherenkov radiation can only occur for particles propagating in a medium, since a Lorentz-invariant vacuum prevents it by energy-momentum conservation \cite{jelley,Macleod:2018zcb}.
However, some Lorentz violating scenarios provide sufficient instabilities so that particles can radiate through the Cherenkov process even in vacuum \cite{Lehnert,Altschul:2017xzx}.
Many aspects about the possibility of vacuum Cherenkov radiation
by Lorentz violating effects have been discussed in the framework of the Standard-Model Extension (SME), within the classical approach to electromagnetic particle radiation \cite{Altschul:2006zz,Altschul:2007tn,Altschul:2007kr,Schober:2015rya,DeCosta:2018nyf}, as well as in the field theory \cite{Kaufhold:2005vj,Kaufhold:2007qd,Colladay:2016rmy,Schreck:2017isa}.
In general, the TeV photons data used to constraint Lorentz violation from the vacuum Cherenkov radiation come from extremely energetic astronomical sources, and therefore even tiny changes in electromagnetic wave propagation can be scrutinized.
 
In this paper we examine the problem of energy loss for a charged particle in vacuum through the Cherenkov effect in the Lifshitz field theory framework  \cite{Horava:2009uw}, establishing an alternative point of view for previous Lorentz symmetry violation studies.
This proposal is mainly motivated by the fact that Lifshitz field theories have a better ultraviolet behavior at the expenses of breaking Lorentz
invariance. This improved behavior is achieved by means of higher spatial derivative terms that are introduced in such a way to avoid the appearance of \emph{ghosts} (negative energy modes), resulting in a theory that exhibits an anisotropic scaling of space and time, i.e. the scaling $x^i\rightarrow \lambda x^i$  whereas $t\rightarrow \lambda^z t$. 
Therefore , this setting is a highly fascinating scenario to examine Lorentz violating effects in phenomenological analyses \cite{Visser:2009fg,Sotiriou:2009gy,Alexandre:2011kr,Mukohyama:2010xz,Nojiri:2010wj,Wang:2017brl}.

The interest of this approach in the description of the problem of computing the rate of radiated energy by a charged particle is threefold: i) the theory is constructed in such a way to have an improved ultraviolet behavior, then we can show that the $z=2$ Lifshitz electrodynamics satisfies the vacuum Cherenkov effect kinematical constraint: the fermion group velocity exceeds the photon velocity for a given value of the three-momentum; ii) this theory also establish physical region in the phase space for this anomalous decay to happen; and iii) this theory admit by construction the definition of only renormalizable interactions, this point will be very important in the computation of the rate of energy loss by a charge in the field theory approach.

Hence, since the time-space asymmetry is a phenomenologically appealing feature to be studied in high energy physics,  we will examine the possibility of the vacuum Cherenkov radiation within the $z=2$ Lifshitz quantum electrodynamics (QED) \cite{Montani:2012ve,Alexandre:2013wua,Bufalo:2015eia,Gomes:2016ixw}.
We start Sec.~\ref{sec2} by reviewing the main aspects of the dynamics for the fermion and gauge fields within the $z=2$ Lifshitz QED.
Moreover, we discuss the general plane wave solutions and coupling for the fields, 
establishing the modified dispersion relations, fermionic energy projection operators, and the photon polarization tensor.
We discuss in Sec.~\ref{sec3} the features of the vacuum process $e^- \to \gamma + e^-$ in the Lifshitz framework.
First, we examine the vacuum Cherenkov kinematical constraint, and show that for a certain a given value of the three-momentum the fermion group velocity exceeds the photon velocity.
Furthermore, we compute the rate of radiated energy for the $z=2$ Lifshitz QED, and discuss the behavior of the instantaneous energy loss and also the emission of soft photons of this decay process.
In Sec.\ref{conc} we summarize the results, and present our final remarks.

%%%%%%%%%%%%%%%%%%%%%%%%%%%%%%%%%%%%
%%%%%%%%%%%%%%%%%%%%%%%%%%%%%%%%%%%%
\section{$z=2$ Lifshitz electrodynamics}
\label{sec2}

We start this section by reviewing and discussing the main points regarding the $z=2$ Lifshitz electrodynamics, more importantly the free field solutions, fermion and gauge fields  dispersion relations and completeness relations  \cite{Alexandre:2013wua,Bufalo:2015eia}.
In particular, the analysis of the dispersion relations is very important since the first constraint upon the vacuum Cherenkov radiation is kinematical, where the fermion group velocity must exceed the photon phase velocity for a certain range of the three-momentum for this anomalous decay to happen.   
We define a gauge and $z=2$ Lifshitz-invariant QED Lagrangian density as
\begin{equation}
\mathcal{L}=\frac{1}{2}F_{0i}F_{0i}-\frac{1}{4}F_{ij}\left(\mu^{2}-\Delta\right)F_{ij}+\overline{\psi}\left(i\gamma_{0}D_{0}-i\mu\gamma_{k}D_{k}-D_{k}D_{k}-m^{2}\right)\psi, \label{eq1}
\end{equation}
where the covariant derivative is $D_{\mu}=\partial_{\mu}+igA_{\mu}$, and the
field strenght is defined as usual $F_{\mu\nu}=\partial_{\mu}A_{\nu}-\partial_{\nu}A_{\mu}$.
The model  \eqref{eq1} is invariant under the $U\left(1\right)$ gauge symmetry
\begin{equation}
\psi\rightarrow e^{i\sigma}\psi,\quad A_{\mu}\rightarrow A_{\mu}+\frac{1}{g}\partial_{\mu}\sigma.
\end{equation}

This theory \eqref{eq1} is known to be super-renormalizable \cite{Alexandre:2013wua}. From the Lagrangian density \eqref{eq1} we may observe that the length dimensions are, in $3+1$ dimensions,
\begin{align}
\left[A_{0}\right] & =\left[\psi\right]=L^{-\frac{3}{2}},\quad\left[A_{i}\right]=L^{-\frac{1}{2}},\cr
\left[g\right] & =L^{-\frac{1}{2}},\quad\left[m\right]=\left[\mu\right]=L^{-1}.
\end{align}

Since we are interested in evaluate the matrix element related to the $e^{-} \rightarrow \gamma+ e^{-}$ process, 
corresponding to the Cherenkov radiation, we need to establish the free field
solutions for the fermionic and gauge field equations.
Hence, in order to construct the solutions for the Dirac fields, we shall consider the following modified free field equation, obtained from \eqref{eq1}, which reads 
\begin{equation}
\left(i\gamma_{0}\partial_{0}-i\mu\gamma_{k}\partial_{k}-\partial_{k}\partial_{k}-m^{2}\right)\psi=0,\label{eq: 0.2}
\end{equation}
 the coefficients of the differential equation are constants, thus
$\psi\left(x\right)=e^{-ipx}\chi\left(p\right)$, will be a solution.
Hence, we obtain 
\begin{equation}
\left(\gamma_{0}p_{0}-\mu\gamma_{k}p_{k}+p^{2}-m^{2}\right)\chi\left(p\right)=0,\label{eq: 0.3}
\end{equation}
 where $p^{2}=p_{i}p_{i}$. For the energy eigenvalues, we have 
\begin{equation}
p_{0}\equiv\pm \mathcal{E}_{p}=\pm\sqrt{\mu^{2}p^{2}+\left(p^{2}-m^{2}\right)^{2}}.\label{eq: 0.4}
\end{equation}
 For each value of $p_{0}$, the solution \eqref{eq: 0.3} has a
two-dimensional solution space. Thus, for the explicit calculation, we choose
\begin{equation}
\chi\left(p\right)=\left(\begin{array}{c}
u_{s}\left(p\right)\\
v_{s}\left(p\right)
\end{array}\right),
\end{equation}
 and we also consider the following representation for the $\gamma$-matrices,
\[
\gamma^{0}=\left(\begin{array}{cc}
\mathbf{1} & 0\\
0 & -\mathbf{1}
\end{array}\right),\quad\vec{\gamma}=\left(\begin{array}{cc}
0 & \vec{\sigma}\\
-\vec{\sigma} & 0
\end{array}\right),
\]
 where $\mathbf{1}$ is a two-dimensional identity matrix, and $\vec{\sigma}$
are the set of Pauli matrices. All these considerations lead to the expressions
\begin{align}
\left(p_{0}+p^{2}-m^{2}\right)u_{s}\left(p\right) & =\mu\left(\vec{\sigma}.\vec{p}\right)v_{s}\left(p\right),\\
\left(p_{0}-p^{2}+m^{2}\right)v_{s}\left(p\right) & =\mu\left(\vec{\sigma}.\vec{p}\right)u_{s}\left(p\right).
\end{align}
 Nevertheless, it is not difficult to show that $u\left(p\right)$
corresponds to those solutions with positive energy, $p_{0}=+\mathcal{E}_{p}$,
whereas $v\left(p\right)$ are solutions with negative energy, $p_{0}=-\mathcal{E}_{p}$.
The next step involves the definition of the energy projection operators,
so that the following relations hold
\begin{align}
\Pi^{+}\left(p\right)u_{s}\left(p\right) & =u_{s}\left(p\right),\\
\Pi^{-}\left(p\right)v_{s}\left(p\right) & =v_{s}\left(p\right),
\end{align}
and also
\begin{equation}
\Pi^{-}\left(p\right)u_{s}\left(p\right)=\Pi^{+}\left(p\right)v_{s}\left(p\right)=0.
\end{equation}
After some algebraic construction, we can show that the operators
\begin{equation}
\Pi^{\pm}\left(p\right)=\frac{\mp\gamma_{0}p_{0}\pm\mu\gamma_{k}p_{k}+p^{2}-m^{2}}{2\left(p^{2}-m^{2}\right)},\label{eq: 0.5}
\end{equation}
 satisfy the above relations, as well as the following identities
\begin{gather}
\Pi^{+}\left(p\right)+\Pi^{-}\left(p\right)=\mathbf{1},\quad\Pi^{+}\left(p\right)\Pi^{-}\left(p\right)=0,\\
\left[\Pi^{\pm}\left(p\right)\right]^{2}=\Pi^{\pm}\left(p\right).
\end{gather}
Furthermore, from the definitions \eqref{eq: 0.5}, we can also show that these solutions satisfy the completeness relations
\begin{align}
\Pi_{\alpha\beta}^{+}\left(p\right) & =\sum_{s=1}^{2}u_{\alpha}\left(p,s\right)\overline{u}_{\beta}\left(p,s\right)=\left[\frac{-\gamma_{0}p_{0}+\mu\gamma_{k}p_{k}+p^{2}-m^{2}}{2\left(p^{2}-m^{2}\right)}\right]_{\alpha\beta},\label{eq: 0.6a}\\
\Pi_{\alpha\beta}^{-}\left(p\right) & =-\sum_{s=1}^{2}v_{\alpha}\left(p,s\right)\overline{v}_{\beta}\left(p,s\right)=\left[\frac{\gamma_{0}p_{0}-\mu\gamma_{k}p_{k}+p^{2}-m^{2}}{2\left(p^{2}-m^{2}\right)}\right]_{\alpha\beta}.\label{eq: 0.6b}
\end{align}
 Finally, taking into account all of these results, we can write the free solutions as the following
\begin{align}
\psi\left(x\right) & =\sum_{r}\int\frac{d^{3}p}{\left(2\pi\right)^{\frac{3}{2}}}\left(\frac{p^{2}-m^{2}}{\mathcal{E}_{p}}\right)^{\frac{1}{2}}\left[b_{r}\left(p\right)u_{r}\left(p\right)e^{-ipx}+d_{r}^{\dagger}\left(p\right)v_{r}\left(p\right)e^{ipx}\right],\label{eq: 0.7a}\\
\overline{\psi}\left(x\right) & =\sum_{r}\int\frac{d^{3}p}{\left(2\pi\right)^{\frac{3}{2}}}\left(\frac{p^{2}-m^{2}}{\mathcal{E}_{p}}\right)^{\frac{1}{2}}\left[b_{r}^{\dagger}\left(p\right)\overline{u}_{r}\left(p\right)e^{ipx}+d_{r}\left(p\right)\overline{v}_{r}\left(p\right)e^{-ipx}\right],\label{eq: 0.7b}
\end{align}
 where, the operators' anti-commutation algebra is given as usual, 
\begin{equation}
\left\{ b_{r}\left(p\right),b_{s}^{\dagger}\left(q\right)\right\} =\left\{ d_{r}\left(p\right),d_{s}^{\dagger}\left(q\right)\right\} =\delta_{rs}\delta\left(\vec{p}-\vec{q}\right).\label{eq: 0.8a}
\end{equation}
 From such construction one can easily show that the equal-time anti-commutation relations is satisfied 
\begin{equation}
\left\{ \psi_{\alpha}(x),\psi_{\beta}^{\dagger}(y)\right\} _{x_{0}=y_{0}}=\delta_{\alpha\beta}\delta^{(3)}\left(\vec{x}-\vec{y}\right).\label{eq: 0.8b}
\end{equation}
 Besides, we can also determine the fermion propagator 
\begin{equation}
S\left(p_{0},p\right)=i\frac{\gamma_{0}p_{0}-\mu\gamma_{k}p_{k}+p^{2}-m^{2}}{p_{0}^{2}-\mu^{2}p^{2}-\left(p^{2}-m^{2}\right)^{2}},\label{eq: 0.09}
\end{equation}
which is in accordance with the free field solutions and operator algebra.
The development for the gauge field follows closely of the fermionic part. First we find the energy eigenvalues, 
\begin{equation}
k_{0}\equiv\pm\Omega_{k}=\pm\sqrt{\mu^{2}k^{2}+k^{4}}.\label{eq: 0.10}
\end{equation}
 In the case of the gauge field, the free field solution is simply
\begin{equation}
A_{\mu}\left(x\right)=\sum_{\lambda}\int\frac{d^{3}k}{\left(2\pi\right)^{\frac{3}{2}}}\left(\frac{1}{2\omega_{k}}\right)^{\frac{1}{2}}\left[a_{i}\left(k\right)\epsilon_{\mu}\left(k,\lambda\right)e^{-ikx}+a_{i}^{\dagger}\left(k\right)\epsilon_{\mu}\left(-k,\lambda\right)e^{ikx}\right].\label{eq: 0.11}
\end{equation}
where the polarization tensor satisfies the normalization condition
\begin{equation}
\eta^{\mu\nu}\epsilon_{\mu}^{*}\left(k,\lambda\right)\epsilon_{\nu}\left(k,\lambda\right)=-1
\end{equation}
and also the completeness relation
\begin{equation}
\sum_{\lambda}\epsilon_{\mu}^{*}\left(k,\lambda\right)\epsilon_{\nu}\left(k,\lambda\right)=g_{\mu\nu} \label{eq: 0.99}
\end{equation}
where $g_{\mu\nu}$ is a `metric' with  well defined components (the difference with the Lorentzian metric $\eta_{\mu\nu}$ is due to the different scaling between time and spatial components). 
Moreover, in order to define the photon propagator and determine the polarization tensor, we must impose
a gauge condition, which is a Lorenz-like condition 
\begin{equation}
\Omega\left[A\right]=\partial_{0}A_{0}-\left(-\Delta+\mu^{2}\right)\partial_{k}A_{k}=0, \label{eq: 1.99}
\end{equation}
 this leads to a non-local gauge-fixing term in the Lagrangian density which, in the Feynman gauge $\left(\alpha=1\right)$,  reads
\begin{equation}
\mathcal{L}_{gf}=-\left(\partial_{0}A_{0}-\left(-\Delta+\mu^{2}\right)\partial_{k}A_{k}\right)\frac{1}{2\left(-\Delta+\mu^{2}\right)}\left(\partial_{0}A_{0}-\left(-\Delta+\mu^{2}\right)\partial_{j}A_{j}\right),\label{eq: 0.12}
\end{equation}
Finally, we see that the nonvanishing components of the gauge field propagator have a well-behaved expression: 
\begin{align}
i\mathcal{D}_{00}\left(k_{0},k\right) & =\frac{\mu^{2}+k^{2}}{k_{0}^{2}-\mu^{2}k^{2}-k^{4}},\label{eq: 0.13a}\\
i\mathcal{D}_{ij}\left(k_{0},k\right) &  =-\frac{\delta_{ij}}{k_{0}^{2}-\mu^{2}k^{2}-k^{4}},\label{eq: 0.13b}
\end{align}
 with no off-diagonal components.
At last, we can determine the metric components $g_{\alpha\beta}$, present in modified completeness relation \eqref{eq: 0.99}, so that they are in agreement with the Feynman propagators \eqref{eq: 0.13a} and \eqref{eq: 0.13b} computed in the gauge \eqref{eq: 1.99}.
This identification yields
\begin{align}  \label{eq: 0.93}
\varepsilon_{0}^{*}\left(k,\lambda\right)\varepsilon_{0}\left(k,\lambda\right) & =\mu^{2}+k^{2},\\
\varepsilon_{0}^{*}\left(k,\lambda\right)\varepsilon_{k}\left(k,\lambda\right) & =0,\\
\varepsilon_{s}^{*}\left(k,\lambda\right)\varepsilon_{l}\left(k,\lambda\right) & =g_{sl}=-\delta_{sl}. \label{eq: 0.95}
\end{align}

These results for the completeness relations of the polarization tensor $\varepsilon_{\mu}\left(k,\lambda\right)$, eq.~ \eqref{eq: 0.99}, as well as for the spinor components $u$ and $v$, eqs.~\eqref{eq: 0.6a} and \eqref{eq: 0.6b}, will be very important in the evaluation of the amplitude related to the decay $e\to e+\gamma$.
 
The last part we need to discuss is about the couplings present in the Lagrangian density \eqref{eq1} and their contribution to the decay process of interest.
One can observe in \eqref{eq1} that the $z=2$ QED presents three vertices: \textit{two three-point vertices}, 
\begin{equation}
\left\langle \overline{\psi}A_{0}\psi \right\rangle \rightarrow ig\gamma_{0},\quad\left\langle \overline{\psi}A_{k}\psi  \right\rangle\rightarrow-ig\left(\gamma_{k}\mu+2p_{k}^{\left(\psi\right)}+p_{k}^{\left(A\right)}\right),
\end{equation}
 which can be conveniently rewritten in a compact form, 
\begin{equation}
\left(\left\langle \overline{\psi}A_{0}\psi \right\rangle,\left\langle \overline{\psi}A_{k}\psi \right\rangle\right)\rightarrow\Lambda_{a}\left(q\right)=ig\left(\gamma_{0},-\gamma_{k}\mu-q_{k}\right),\label{eq: 0.14}
\end{equation}
 where we have introduced, by means of notation, the
index $a=0,...,3$, which should not be confused with spacetime index,
and $q_{k}\equiv2p_{k}^{\left(\psi\right)}+p_{k}^{\left(A\right)}$;
and one \textit{four-point vertex}, $\left\langle \overline{\psi}\psi A_{i}A_{j} \right\rangle\rightarrow-2ig\delta_{ij}$.
But since we are interested in the decay $e\to e+\gamma$, we shall consider only the three-point vertices \eqref{eq: 0.14}.
 
In the next section we shall analyze the Cherenkov radiation in the Lorentz violating framework of the Lifshitz electrodynamics.
The interest in this kind of radiation is due its unique signature of Lorentz violation.
We start by discussing the kinematical constraint related with the fermionic group velocity and the photon phase velocity; this analysis establishes the threshold of the three-momentum where the radiation can occur.

Furthermore, after verifying that the Cherenkov kinematical constraint is satisfied  by the $z=2$ Lifshitz electrodynamics, we proceed to the evaluation of the rate of energy loss through vacuum Cherenkov radiation.
The rate of radiated energy is strongly dependent on the cutoff due to energy-momentum conservation  \cite{Altschul:2007tn}, meaning that two types of processes can occur based on the energies of the photons involved: i) the instantaneous rate of emission, in which the charge emits a single energetic photon, drops below the Cherenkov threshold, and stops emitting, and ii) the emission of very soft photons, where the particle's energy is not lowered below the threshold, and so the charge will continue to radiate afterwards the emission. 
We shall mainly discuss the instantaneous emission of photons, but will present some general remarks about the emission of very soft photons in the Lifshitz framework.

%%%%%%%%%%%%%%%%%%%%%%%%%%%%%%%%%%%%%%%%%%%%%%%%%%%%
%%%%%%%%%%%%%%%%%%%%%%%%%%%%%%%%%%%%%%%%%%%%%%%%%%%%
\section{Vacuum Cherenkov radiation in $z=2$ Lifshitz electrodynamics}
\label{sec3}

\subsection{Kinematical constraint}

Before we start our analysis on the rate of radiated energy of the Cherenkov process in the Lifshitz electrodynamics, we must be sure that the kinematical constraint upon this decay is fulfilled, allowing this to occur in this Lorentz violating framework.
First, we can recast the dispersion relations by a rescaling $(p_0, k_{0})\to\mu (p_0, k_{0})$, yielding 
\begin{align}
  \Omega_{k}&=\sqrt{k^{2}+\frac{k^{4}}{\mu^{2}}}, \label{1.10a}\\
  \mathcal{E}_{p}&=\sqrt{\frac{1}{\mu^{2}}p^{4}+\left(1-\eta\right)p^{2}+m_{r}^{2}},\label{1.10}
\end{align}
with the mass $m_{r}^{2}=\frac{1}{\mu^{2}}m^{4}$ and $\eta=\frac{2 m_r}{\mu }$. In this form, we see that the usual relativistic dispersion relations are recovered as  $\eta=0$ and $\mu\to\infty$.
The free parameter $\mu^2$ can be chosen as: the electron mass $m_r \simeq 0,5\,{\rm MeV}$ and the GUT scale $\mu \simeq 10^{16}\,{\rm GeV}$, which results into $\eta \simeq 10^{-19}$; these values  are within the Lorentz symmetry violation bounds \cite{Alexandre:2013wua}. 
%{\color{red} In what follows, we shall consider $\eta =0$ in our analysis, and focus on the main lore of Lifshitz field theory that is the addition of (spatial) higher derivative terms enhancing renormalizability without jeopardizing unitarity.\footnote{{\color{red}In fact, the choice $\eta =0$ in the dispersion relation \eqref{1.10} can be understood as a simple change $p \to p/\sqrt{1-\eta}$, implying that $\mathcal{E}_{p}=\sqrt{\frac{1}{\bar{\mu}^{2}}p^{4}+p^{2}+m_{r}^{2}}$, where $\bar{\mu}^{2} = \mu^{2}\left(1-\eta\right)^2\simeq \mu^{2}$ under the consideration that $\eta \simeq 10^{-19}$. Furthermore, this new form of the dispersion relation shows that it depends actually only on two parameters $\bar{\mu}$ (or $\mu$) and $m_r$, and not three as \eqref{1.10} could naively suggest.} }  }

It is well known that Cherenkov radiation is possible only if there is a range of three-momentum for which the fermion group velocity exceeds the photon phase velocity in medium materials \cite{jelley}, but also for Lorentz-violating vacua \cite{Altschul:2006zz,Altschul:2007tn,Altschul:2007kr}.
In our model we have that 
\begin{align}
v_{\rm g} & =\frac{d\mathcal{E}_{p}}{dp}  =\frac{\frac{2}{\mu^{2}}p^{3}+\left(1-\eta\right)p}{\sqrt{\frac{1}{\mu^{2}}p^{4}+\left(1-\eta\right)p^{2}+m_{r}^{2}}}
\end{align}
and
\begin{align}
v_{\rm ph} & =\frac{\Omega_{k}}{k} =\sqrt{1+\frac{k^{2}}{\mu^{2}}} .
\end{align}
By a simple numerical analysis, we can check that for values above $p>p_{\rm min}=3,16228\times10^{3}~{\rm TeV}$
we have $v_{g}>v_{ph}$.
Hence, as long as we are in the region $p>p_{\rm min}$, vacuum Cherenkov radiation can occur in the $z=2$ Lifshitz QED.
Actually, in our discussion of the phase space related to this decay, we will find a tighter bound upon the charge three-momentum by demanding the reality of the physical region.
Now that we have established the momentum threshold, and consequently the physical region of interest, we can proceed to the computation of the instantaneous rate of radiate energy for the $e\to e+\gamma$ in the $z=2$ Lifshitz QED.   

\subsection{Rate of radiated energy}

Since we are interested in computing the energy loss associated with the vacuum Cherenkov scattering in the $z=2$ Lifshitz electrodynamics scenario, consisting in a decay process  $1\rightarrow2+3$, let us review some general results necessary for the development of this analysis \cite{Kaufhold:2005vj,Kaufhold:2007qd}.
The energy-momentum loss of a Lorentz invariant charged particle per unit of time is equal to the photon four-momentum $k$ weighted by the scattering amplitude squared and integrated over phase space,
\begin{equation}
\frac{dp^{\mu}}{dt}=\int Dk\left|\mathcal{M}\right|^{2}k^{\mu}
\end{equation}
where $Dk$ is the phase-space invariant measure.
The rate of total radiated energy is obtained from the time
component of the above expression.

\begin{figure}[t]
\vspace{-0.5cm}
\includegraphics[height=5.7\baselineskip]{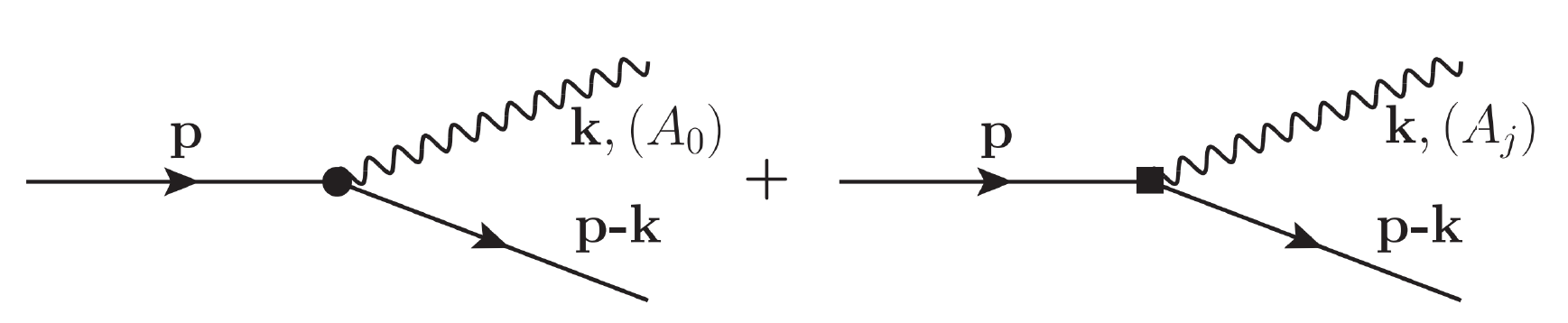}
\centering\caption{Tree-level Feynman graphs for vacuum Cherenkov radiation in the $z=2$ Lifshitz electrodynamics, the left panel corresponds to the coupling $\left\langle \overline{\psi}A_{0}\psi\right\rangle$ while the right one to the coupling $\left\langle \overline{\psi}A_{j}\psi\right\rangle$.}
\label{treelevel}
\end{figure}

In the presence of Lorentz violating effects, however, the above expression is generally no longer valid, because such effects also modify the energy-momentum tensor for the particle, and hence the energy of the particle is not necessarily equal to the time component of its four-vector momentum.
On the other hand, since the emission rate is identically zero without the Lorentz violation, only the desired term survives at leading $\mu^2$ order.
Thus, the rate of radiated energy, in the first-order in $\mu^2$, can be expressed as
\begin{equation}
W\approx-\dot{p}_{0} = \int k_{0}d\Gamma, \label{eq100}
\end{equation}
where $d\Gamma$ is the differential decay rate for the given process \cite{mandl_shaw}.
The contributions to the Cherenkov radiation process at tree-level are represented by the Feynman diagrams in Fig.~\ref{treelevel}.

The differential decay rate of our interest can readily be obtained by identifying it with the time derivative of the probability for the decay $1\rightarrow2+3$, i.e. $d\Gamma = dP/dt$, where 
$P=\frac{\left|\left\langle f|i\right\rangle \right|^{2}}{\left\langle f|f\right\rangle \left\langle i|i\right\rangle }$.  Hence, for the process $e^{-} \rightarrow \gamma + e^{-}$ the differential decay rate is explicitly written as
\begin{equation} 
d\Gamma=\frac{m_r}{\mathcal{E}_{p_{i}}}\frac{d^{3}k_{f}}{\left(2\pi\right)^{3}2\Omega_{k_{f}}}\frac{m_r d^{3}p_{f}}{\left(2\pi\right)^{3}\mathcal{E}_{p_{f}}}\left(2\pi\right)^{4}\delta^{4}\left(p_{i}-p_{f}-k_{f}\right)\left|\mathfrak{M}\right|^{2}
\label{eq101}
\end{equation}
where the normalization factors are chosen accordingly for bosonic
and fermionic fields, whereas $\Omega_{k_{f}}$ and $\mathcal{E}_p$ are the dispersion relations \eqref{1.10a} and \eqref{1.10}, respectively, and $m_r ^2=m^{4}/\mu^{2}$ is the fermionic mass.
Thus, substituting the result \eqref{eq101} into the expression \eqref{eq100}, we can write the rate of radiated energy as the following
\begin{equation}
W=\frac{\mu^{2}}{8\pi^{2}E_{p_{i}}}\int\frac{d^{3}k_{f}d^{3}p_{f}}{E_{p_{f}}}\delta^{4}\left(p_{i}-p_{f}-k_{f}\right)\left|\mathfrak{M}\right|^{2} \label{eq102}
\end{equation}
Furthermore, by calculation purposes and to make the reaction kinematics visible, it is convenient to express the integration over the variables $p_{f}$ as
\begin{equation}
\int\frac{d^{3}p_{f}}{2\mathcal{E}_{p_{f}}}=\int d^{4}p_{f}\delta\left(p_{f}^{02}-\mathcal{E}_{p_{f}}^{2}\right)\theta\left(p_{f}^{0}\right)
\end{equation}
which allow us to rewrite Eq.~\eqref{eq102}  in a convenient form for the remaining analysis
\begin{align}\label{eq103}
W & =\frac{m_{r}^{2}}{4\pi^{2}\mathcal{E}_{p_{i}}}\int d^{3}k\delta\left(\left(\bar{p}-\bar{k}\right)^{2}-\mathcal{M}_{p-k}^{2}\right)\theta\left(\mathcal{E}_{p}-\Omega_{k}\right)\left|\mathfrak{M}\right|^{2}
\end{align}
where we have defined the notation
\begin{equation}
p^{02}-\mathcal{E}_{p}^{2}=\bar{p}^{2}-\left(\frac{1}{\mu^{2}}p^{4}-\eta p^{2}+m_{r}^{2}\right)\equiv\bar{p}^{2}-\mathcal{M}_{p}^{2}
\end{equation}
as matter of simplification, since we are using the bar notation $\bar{p}$ for a  ``four-dimensional'' momentum, also we have rewritten the momenta as $k_{f}=k$ and $p_{i}=p$.

Once we have concluded the formal development of our analysis, by establishing an expression for the rate of radiated energy \eqref{eq103}, we should proceed now to the computation of the scattering matrix $\mathfrak{M}$ in \eqref{eq103} due to the $z=2$ Lifshitz couplings \eqref{eq: 0.14}. 
Hence, the corresponding element of the matrix $\mathfrak{M}$ for the $e^{-} \rightarrow \gamma + e^{-}$ process, depicted in Fig.~\ref{treelevel}, reads
\begin{align}
\mathfrak{M} & =ig\overline{u}\left(p_{f},s_{f}\right)\Lambda\left(k_{f},\lambda\right)u\left(p_{i},s_{i}\right)  \label{eq104}
\end{align}
where we have defined the dependence with the polarization tensor components as
\begin{align}
\Lambda=ig\gamma_{0}\varepsilon_{0}-ig\varepsilon_{k}\left(\gamma_{k}\mu-2\left(p_{i}\right)_{k}+\left(k_{f}\right)_{k}\right)
 \end{align}
 
Since the electrons are not polarized, in order to compute the squared scattering matrix in Eq.~\eqref{eq103}, we average over initial spins
and sum over final spins states, that is $\left|\mathfrak{M}\right|^{2}\rightarrow\frac{1}{2}\sum_{\rm spin}\left|\mathfrak{M}\right|^{2}$.
Hence, we can make use of the completeness relations eqs.~\eqref{eq: 0.6a} and \eqref{eq: 0.6b}, so that it yields
\begin{align} \label{eq105}
\frac{1}{2}\sum_{spin}\left|\mathfrak{M}\right|^{2}  & =\frac{g^{2}}{8\left(\left(p-k\right)^{2}-m^{2}\right)\left(p^{2}-m^{2}\right)}\cr
 & \times {\rm Tr}\Big[\left(-\gamma_{0}\left(p_{0}-k_{0}\right)+\mu\gamma_{k}\left(p_{k}-k_{k}\right)+\left(p-k\right)^{2}-m^{2}\right)\Lambda\left(k,\lambda^{*}\right) \cr
 &\times \left(-\gamma_{0}p_{0}+\mu\gamma_{j}p_{j}+p^{2}-m^{2}\right)\Lambda\left(k,\lambda\right)\Big]
\end{align}
Furthermore, we can compute the trace over the Dirac matrices in \eqref{eq105}, use the polarization states \eqref{eq: 0.93}--\eqref{eq: 0.95}, and then work the algebraic part of the process kinematics, to find the expression
\begin{align} \label{eq106}
\frac{1}{2}\sum_{spin}\left|\mathfrak{M}\right|^{2} & =2g^{2}\left[-p^{2}+pk\cos\theta\right] \cr
&+\frac{g^{2}}{\left(p^{2}-2pk\cos\theta+k^{2}-\mu m_{r}\right)\left(p^{2}-\mu m_{r}\right)}\biggl\{\left(\mathcal{E}_{p}-\varOmega_{k}\right)\mathcal{E}_{p}\left[\mu^{2}+2p^{2}-2pk\cos\theta\right]\cr
 & -\mu^{2}\bigg[8p^{4}-16p^{3}k\cos\theta+4p^{2}k^{2}\cos^{2}\theta+3k^{2}p^{2}-k^{3}p\cos\theta  \bigg] \cr
 & -\mu^{2}\bigg[-\mu m_{r}\left(8p^{2}-8pk\cos\theta+k^{2}\right)+2\mu^{2}m_{r}^{2}\bigg]\biggr\} .
\end{align}
where $\theta$ is chosen as the opening angle between the incoming electron and the outgoing photon, and it is determined from the energy-momentum conservation given by the delta function in \eqref{eq103}.
The analysis of the kinematics of the decay is also crucial to establish the physical phase space where the decay can happen, which corresponds to the allowed integration interval in the momentum integral in \eqref{eq103}.
Under these considerations, we shall work with the variables $d^{3}k=2\pi d\cos\theta~ |\vec{k}| ^2 d|\vec{k}|$.
Hence,  the energy-momentum conservation for the decay process $\left(\bar{p}-\bar{k}\right)^{2} = \mathcal{M}_{p-k}^{2}$, implies the following relation
\begin{align} \label{eq310}
2 \mathcal{E}_p \Omega_k-2\left\{ \left(1-\eta\right)pk+\frac{2}{\mu^{2}}\left(p^{3}k-p^{2}k^{2}+pk^{3}\right)\right\} \cos\theta+\frac{1}{\mu^{2}}2p^{2}k^{2}-\eta k^{2} & =0,
\end{align}
where $k\equiv |\vec{k}|$.
This energy balance equation can be used to arrive at the radiation condition for the $e^-\rightarrow e^-+\gamma$ process within the Lifshitz framework  \cite{jelley}.
After some algebraic manipulations, we find that the opening angle $\theta$, in the regime $ \mu^2 >  (p^2,k^2) $, is cast as 
\begin{equation}
\cos\theta \approx 1 +\frac{\eta (p-k)}{2 p}-\frac{3}{2\mu^{2}}  (p-k)^2 +\mathcal{O} \left( \frac{(p-k)^4}{\mu^4}\right).
\end{equation}
It is straightforward to recognize that the removal of the Lorentz violating effects through $\mu^2\to \infty$ , implies that $\cos\theta=1$, which reflects the fact that the process $e^-\rightarrow e^-+\gamma$ has a vanishing radiation rate in the Lorentz invariant QED.
On the other hand, for finite values of the parameters $\mu^2\neq 0$ and $\eta\neq 0$, one can realize that there is a region in the phase space where $\cos\theta <1$, even in vacuum.
This is a very interesting result, since this radiation condition corresponds to the availability of a physical phase space for the anomalous decay, and it also corroborates the kinetic condition discussed above, where we have established the existence of a physical region for the decay corresponding to $p>p_{\rm min}$ (for smaller momentum values, the radiation rate is strictly zero).

Furthermore, since the energy conservation requires that the allowed values for $k$ are such that the relation \eqref{eq310} is satisfied for a given value of $\theta$, thus the integration over $\theta$ restricts the region of integration over $k$.
To determine this constraint upon $k$, it is worth to rewrite the integration in $k$ as the following
\begin{align}
\mathcal{T} &= \int k^2 dk \theta\left(\mathcal{E}_{p}-\Omega_{k}\right) \int_{-1}^{+1}d\cos \theta\delta\left(\left(\bar{p}-\bar{k}\right)^{2}-\mathcal{M}_{p-k}^{2}\right) \frac{1}{2}\sum_{spin}\left|\mathfrak{M}\right|^{2}  \cr
&= \int k^{2}dk \theta\left(\mathcal{E}_{p}-\Omega_{k}\right) \frac{1}{2\left\{ \left(1-\eta\right)pk+\frac{2}{\mu^{2}}\left(p^{3}k-p^{2}k^{2}+pk^{3}\right)\right\} }  \cr
 & \times \int_{-1}^{+1}d\cos\theta\delta\left(\cos\theta-\frac{\frac{1}{\mu^{2}}2p^{2}k^{2}-\eta k^{2}+2\mathcal{E}_{p}\varOmega_{k}}{2\left\{ \left(1-\eta\right)pk+\frac{2}{\mu^{2}}\left(p^{3}k-p^{2}k^{2}+pk^{3}\right)\right\} }\right)\frac{1}{2}\sum_{spin}\left|\mathfrak{M}\right|^{2}  \label{eq110}
\end{align}
Finally, from the expression \eqref{eq110} we can conclude that the condition $\cos\theta \in [-1,1]$ restricts the magnitude of the photon momentum $k$ to the values\footnote{ In fact, eq.\eqref{eq310} gives four nonvanishing roots, two of them are imaginary for any value of $p$, while the remaining two result into in eqs.\eqref{eq110a} and \eqref{eq110b}.  }
\begin{align} 
& k_{+}  =-\frac{\gamma\mu^{4}}{64\xi^{2}}+\frac{1}{2}\sqrt{\beta}\cr
&+\frac{1}{2}\sqrt{\frac{3\gamma^{2}\mu^{8}}{1024\xi^{4}}-\frac{\delta\mu^{2}}{8\xi^{2}}+\frac{\gamma\mu^{6}}{256\xi^{4}\sqrt{\beta}}\left(\delta+\frac{\gamma^{2}\mu^{6}}{64\xi^{2}}\right)+\frac{\mu^{3}}{2\xi\sqrt{\beta}}\left(12\xi^{4}-8\eta \xi^{2}+6\xi^{2}+\eta^{2}-\eta\right)-\beta}\label{eq110a}\\
 & k_{-}= -\frac{\gamma\mu^{4}}{64\xi^{2}}+\frac{1}{2}\sqrt{\beta}\cr
 &-\frac{1}{2}\sqrt{\frac{3\gamma^{2}\mu^{8}}{1024\xi^{4}}-\frac{\delta\mu^{2}}{8\xi^{2}}+\frac{\gamma\mu^{6}}{256\xi^{4}\sqrt{\beta}}\left(\delta+\frac{\gamma^{2}\mu^{6}}{64\xi^{2}}\right)+\frac{\mu^{3}}{2\xi\sqrt{\beta}}\left(12\xi^{4}-8\eta\xi^{2}+6\xi^{2}+\eta^{2}-\eta\right)-\beta} \label{eq110b}    
 \end{align} 
where we have introduced the factors $\alpha$, $\beta$, $\gamma$ and $\delta$ 
for simplicity of notation, and define them in the Appendix \ref{apA}.  Moreover, we have also introduced the parameter $\xi= \frac{p}{\mu}$, which will work as the perturbative parameter in the analysis of the approximated expression for the radiation rate.

Hence, the above discussion about  availability of the physical phase space for the electron decay, displayed by Eqs.~\eqref{eq110a} and \eqref{eq110b}, permit us to rewrite \eqref{eq103} as the instantaneous rate
\begin{align}
W_{\rm inst} & =\frac{m_{r}^{2}}{4\pi \mathcal{E}_{p}} \int_{k_{-}}^{k_{+}} \frac{k^{2}dk }{  \left(1-\eta\right)pk+\frac{2}{\mu^{2}}\left(p^{3}k-p^{2}k^{2}+pk^{3}\right)  }\cr
 & \times \int_{-1}^{+1}d\cos\theta\delta\left(\cos\theta-\frac{\frac{1}{\mu^{2}}2p^{2}k^{2}-\eta k^{2}+2\mathcal{E}_{p}\varOmega_{k}}{2\left\{ \left(1-\eta\right)pk+\frac{2}{\mu^{2}}\left(p^{3}k-p^{2}k^{2}+pk^{3}\right)\right\} }\right)\frac{1}{2}\sum_{spin}\left|\mathfrak{M}\right|^{2}
 \label{eq111}
\end{align}
Finally, substituting the expression for the scattering matrix \eqref{eq106} into  \eqref{eq111}, we can perform the integrations over $\theta$ and over the momentum $k$ with the use of Mathematica.
However, the complete result is not enlightening, but we can discuss some asymptotic regions of interest, in particular $\xi< 1$ and $\xi>1$.
It is straightforward to conclude that the case of $\xi>1$ corresponds to a region where this effective theory stops working, hence it is not of interest.
Hence, we can conclude that the only region of physical interest for the Cherenkov decay is for $\xi<1$, more precisely this anomalous decay can happen in the Lifshitz QED only in the region $p_{\rm min}<p<\mu$.
This discussion implies that the instantaneous rate read
\begin{align}
W_{\rm inst} & =0,\quad p<p_{\rm min}\\
W_{\rm inst} & \approx  \frac{g^{2}}{2560\pi} \sqrt{\frac{3}{2}}\left[1546+2391\eta\right]\frac{m_{r}^{6}}{\mu^{2}\xi^{9}}  +\mathcal{O}\left( \frac{g^2 m_{r}^{6}}{\mu^{2}\xi^7}\right),\quad p_{\rm min}<p<\mu  \label{eq115}
\end{align}
We observe that at large energies the instantaneous rate $W_{\rm inst}$ is a rapidly decreasing function of $\xi$.
Moreover, as we would expect, the limit $\mu   \rightarrow \infty $ of eq. \eqref{eq115}   gives a vanishing result to the Cherenkov radiation.

As previously discussed, in addition to the instantaneous energy loss, there is the possibility of the emission of lower-energy photons with energies $k< k^-$.
In this regime, the charged particle continue to radiate after the emission of one of these photons because its energy is still above the threshold, which makes it reasonable to approximate the energy losses for these photons as a continuous process \cite{Altschul:2007tn}.
Actually, the instantaneous rate $W_{\rm inst} = \int_{k_{-}}^{k_{+}} W (k) dk $ can be understood as the rate of photon emission per unit energy $W (k) = \mathcal{P} (k) / \Omega_k$, i.e. the charged particle radiates a single high-energy photon.
Within this interpretation, the continuous rate of radiating power related with soft photons can be written as $\mathcal{P}_{\rm soft} = \int_{0}^{k_{-}} \mathcal{P} (k) dk $ and we can use the previous results derived for the evaluation of the instantaneous rate, in particular eq.~\eqref{eq111}, to compute this expression.  
Hence, in the region $p_{\rm min}<p<\mu $, we find that the radiating power related with the emission of soft photons is
\begin{equation}\label{eq116}
\mathcal{P}_{soft} \approx \frac{3g^{2}}{40960\pi}\left[2\left(796\sqrt{6}-1155\right)+\left(2072\sqrt{6}-3585\right)\eta\right]\frac{m_{r}^{7}}{\mu^{2}\xi^{10}} +\mathcal{O}\left( \frac{g^2 m_{r}^{7}}{\mu^{2}\xi^8}\right).
\end{equation}
Moreover, we observe that the ratio of instantaneous by the soft emission is greater than unity for a finite value of the parameters, that is, we have $W_{inst}/\left(\mathcal{P}_{soft}\xi/m_{r}\right) \approx\frac{6184\sqrt{\frac{2}{3}}}{796\sqrt{6}-1155} >1$. This shows that although the two types of losses are comparable in this energy regime (since the ratio is only slightly greater than the unit), the instantaneous emission still is more important over the emission of very soft photons in the $z=2$ Lifshitz framework.

%%%%%%%%%%%%%%%%%%%%%%%%%%%%%%%%%%%%%%%%%%%%%%%%%%%%%%%%%%%%%%%%%%%%%%%
\section{Final remarks}
\label{conc}

In this paper, we have studied vacuum effects of a Lorentz violating quantum electrodynamics in the context of $z=2$ Lifshitz field theory.
Anomalous decay processes are a suitable scenario to study due its unique signature of Lorentz violation.
In particular, we analyzed the rate of radiated energy from a charged particle through vacuum Cherenkov radiation.

We started our analysis of vacuum Cherenkov radiation described in the $z=2$ Lifshitz framework by considering the instantaneous rate of radiated energy for the process $e^{-} \rightarrow \gamma+ e^{-}$, where the charge emits a single energetic photon, drops below the Cherenkov threshold, and stops emitting.
In the instantaneous case we observed that it only occurs in the regime $p_{\rm min}<p<\mu $, and  the emission rate at large energies behaves as a decreasing function of $\xi $ ($W_{\rm inst}\sim \xi^{-9}$).
In addition to the instantaneous emission, emission of very soft photons may also occur, where the charge's energy is not lowered below the threshold, and thus the charge continue to radiate afterwards. 
However, after computing these radiation rates, we found that $W_{\rm inst}/W_{\rm soft}  >1$ in the high-energy regime, revealing  the importance of instantaneous emission face to the soft emission of soft photons in the case of $z=2$ Lifshitz electrodynamics.

It is necessary to remark that the $z=2$ Lifshitz contributions to the vacuum processes are highly nontrivial.
We emphasize that the nonvanishing radiated energy rate has two underlying causes: it is the severe departure of the dispersion relations of the electron and photon fields in the $z=2$ Lifshitz electrodynamics in relation to the usual QED, with higher spatial derivative contributions, that causes this theory to fulfill the kinematical constraint, allowing this anomalous decay to happen, but also the unique couplings contributions in the evaluation of the decay amplitude. 

Based on the interesting outcome that the $z=2$ Lifshitz effects have in the vacuum decay processes, in particular in the emission of soft photons, we believe that the signatures of Lorentz violation 
in the Bremmstrahlung contributions to the tree level Coulomb scattering 
 deserve further analysis and discussion, analysing whether the Bremmstrahlung cancels the radiate corrections, as it happens in usual QED.

%%%%%%%%%%%%%%%%%%%%%%%%%%%%%%%%%%%%%%%%%%%%%%%%%%%%%%%%%%%%%%%%%%%%%%%
%%%%%%%%%%%%%%%%%%%%%%%%%%%%%%%%%%%%%%%%%%%%%%%%%%%%%%%%%%%%%%%%%%%%%%%
 \subsection*{Acknowledgements}

The authors would like to thank the anonymous referee for his/her comments and suggestions to improve this paper.
R.B. acknowledges partial support from Conselho
Nacional de Desenvolvimento Cient\'ifico e Tecnol\'ogico (CNPq Projects No. 305427/2019-9 and No. 421886/2018-8) and Funda\c{c}\~ao de
Amparo \`a Pesquisa do Estado de Minas Gerais (FAPEMIG Project No. APQ-01142-17).

 %%%%%%%%%%%%%%%%%%%%%%%%%%%%%%%%%%%%%%%%%%%%%%%%%%%%%%%%%%%%%%%%%%%%%%%%%%%%%%%%%%
\appendix 

\section{Complementary expressions}
\label{apA}

In this appendix we present by complementarity some auxiliary lengthy coefficient expressions, $\alpha$, $\beta$, $\gamma$ and $\delta$, present in the photon momentum roots, eqs.\eqref{eq110a} and \eqref{eq110b},
\begin{align}
\beta=&\frac{m_{r}^{4}}{24\ 2^{2/3}\xi^{4}\alpha}-\frac{2\mu^{2}\sqrt[3]{2}m_{r}^{2}}{3\alpha}+\frac{\mu^{2}\eta m_{r}^{2}}{2\ 2^{2/3}\xi^{2}\alpha}-\frac{9\mu^{2}m_{r}^{2}}{4\ 2^{2/3}\xi^{2}\alpha}-\frac{\mu^{2}\eta^{2}m_{r}^{2}}{48\ 2^{2/3}\xi^{4}\alpha}+\frac{\mu^{2}\eta}{4}-\frac{\mu^{2}}{2}+\frac{\mu^{2}\eta^{2}}{48\xi^{2}} \cr
&+\frac{m_{r}^{2}}{6\xi^{2}}-\frac{5\mu^{2}\xi^{2}}{12}+\frac{\alpha}{3\sqrt[3]{2}}+\frac{\sqrt[3]{2}\mu^{4}\xi^{4}}{3\alpha}-\frac{\mu^{4}\eta\xi^{2}}{2^{2/3}\alpha}+\frac{\mu^{4}\xi^{2}}{2^{2/3}\alpha}+\frac{\mu^{4}\eta^{2}}{3\ 2^{2/3}\alpha}-\frac{\mu^{4}\eta}{2\ 2^{2/3}\alpha}+\frac{3\mu^{4}}{8\ 2^{2/3}\alpha}\cr
&+\frac{\mu^{4}\eta^{3}}{8\ 2^{2/3}\xi^{2}\alpha}-\frac{3\mu^{4}\eta^{2}}{16\ 2^{2/3}\xi^{2}\alpha}+\frac{\mu^{4}\eta^{4}}{384\ 2^{2/3}\xi^{4}\alpha} \\
\alpha_{1}=&-\frac{m_{r}^{6}}{32\xi^{6}}+\frac{3\mu^{2}m_{r}^{4}}{2\xi^{2}}-\frac{9\mu^{2}\eta m_{r}^{4}}{16\xi^{4}}-\frac{135\mu^{2}m_{r}^{4}}{32\xi^{4}}+\frac{3\mu^{2}\eta^{2}m_{r}^{4}}{128\xi^{6}}+\frac{57\mu^{4}\xi^{2}m_{r}^{2}}{4} \cr
&-\frac{135\mu^{4}\eta m_{r}^{2}}{8}+\frac{207\mu^{4}m_{r}^{2}}{8}+\frac{69\mu^{4}\eta^{2}m_{r}^{2}}{16\xi^{2}}-\frac{45\mu^{4}\eta m_{r}^{2}}{4\xi^{2}}+\frac{405\mu^{4}m_{r}^{2}}{32\xi^{2}}-\frac{27\mu^{4}\eta^{2}m_{r}^{2}}{64\xi^{4}}-\frac{3\mu^{4}\eta^{4}m_{r}^{2}}{512\xi^{6}} \cr
& +2\mu^{6}\xi^{6}-\frac{9\mu^{6}\eta\xi^{4}}{2}+\frac{9\mu^{6}\xi^{4}}{2}+\frac{51\mu^{6}\eta^{2}\xi^{2}}{16}-\frac{45\mu^{6}\eta\xi^{2}}{8}+\frac{27\mu^{6}\xi^{2}}{8}+\frac{33\mu^{6}\eta^{4}}{64\xi^{2}} \cr
&-\frac{45\mu^{6}\eta^{3}}{32\xi^{2}}+\frac{135\mu^{6}\eta^{2}}{128\xi^{2}}-\frac{45\mu^{6}\eta^{3}}{32}+\frac{63\mu^{6}\eta^{2}}{32}-\frac{27\mu^{6}\eta}{16}+\frac{27\mu^{6}}{32}+\frac{9\mu^{6}\eta^{5}}{256\xi^{4}}-\frac{27\mu^{6}\eta^{4}}{512\xi^{4}}+\frac{\mu^{6}\eta^{6}}{2048\xi^{6}}  \\
\alpha_{2}=&\frac{3\mu^{4}}{8\xi^{2}}\left(40\xi^{6}-28\eta\xi^{4}+12\xi^{4}+6\eta^{2}\xi^{2}-4\eta\xi^{2}-\eta^{3}+\eta^{2}+\frac{8m_{r}^{2}}{\mu^{2}}\right)\cr
&-\frac{\mu^{4}}{256\xi^{4}}\left(-64\xi^{4}+24\eta\xi^{2}-12\xi^{2}-\eta^{2}+\frac{4m_{r}^{2}}{\mu^{2}}\right)^{2} \\
\gamma=&\frac{8\xi}{\mu^{3}}\left(6\xi^{2}-\eta\right)\\
\delta=& 64\xi^{4}-24\eta\xi^{2}+12\xi^{2}-\frac{4m_{r}^{2}}{\mu^{2}}+\eta^{2}
\end{align}
 
 %%%%%%%%%%%%%%%%%%%%%%%%%%%%%%%%%%%%%%%%%%%%%%%%%%%%%%%%%%%%%%%%%%%%%%%%%%%%%%%%%%%%%%%%%%%%%%%%%%%%%%%%%%%%%%%%%

\global\long\def\link#1#2{\href{http://eudml.org/#1}{#2}}
 \global\long\def\doi#1#2{\href{http://dx.doi.org/#1}{#2}}
 \global\long\def\arXiv#1#2{\href{http://arxiv.org/abs/#1}{arXiv:#1 [#2]}}
 \global\long\def\arXivOld#1{\href{http://arxiv.org/abs/#1}{arXiv:#1}}

%%%%%%%%%%%%%%%%%%%%%%%%%%%%%%%%%%%%%%%%%%%%%%%%%%%%%%%%%%%%%%%%%%%%%%%%%%%%%%%%%%%%%%%%%%%%%%%%%%%%%%%%%%%

\end{document}